\documentstyle[12pt]{article}

\begin{document}

\begin{flushright}
OCHA-PP-77\\
March 1996
\end{flushright}
\vspace{1cm}
\begin{center}
{\Large\bf BEING FASCINATED BY STRINGS AND MEMBRANES: IS 
KIKKAWA-TYPE PHYSICS POSSIBLE AT OCHANOMIZU?}\\
\vspace{1cm}
A. SUGAMOTO\\
\it Department of Physics, Ochanomizu University, \\
2-1-1, Otsuka, Bunkyo-ku, Tokyo, 112 Japan 
\end{center}
\vspace{1cm}

\begin{abstract}
On the occasion of the 60th birthday of Professor 
Keiji Kikkawa, Kikkawa-type physics performed at Ochanomizu was 
personally reviewed, and the generation of the metric is discussed with 
the condensation of the string fields.
\end{abstract}
\vspace{0.5cm}

\section{Personal Memories with Professor Keiji Kikkawa and the 
Kikkawa-Type Physics at Ochanomizu}
It is my great pleasure to contribute to the proceedings of the workshop 
held at Osaka to celebrate the 60th birthday of Professor Keiji Kikkawa.
I am very much influenced by his physics, especially by his papers on 1) 
the light-cone field theory of string (Its Japanese version included in 
Soryushiron Kenkyu was my favorite.), 2) his lecture note on the 
superstrings given just before the string fever started (I think everybody 
should begin with this lecture note when he or she wants to do something 
in strings.), 3) hadronic strings with quarks at the ends, and 4) the path 
integral formulation of the Nambu-Jona-Lasinio model. Personally, 
Professor Kikkawa cited my paper on the dual transformation in gauge 
theories at the Tokyo conference in 1978, without which I could not have 
survived in our particle physics community and would definitely be 
engaged in another job now. Therefore I am greatly indebted to him for his 
guidance in physics. It was probably 1979 summer when I went abroad for 
the first time with my late friend Dr. Osamu Sawada, and we stayed at 
Professor Hirotaka Sugawara's residence in Honolulu. The topical 
conference was held at the moment, and Kikkawa-san came to Honolulu to 
attend it. Kikkawa-san, Sugawara-san, Sandip Pakvasa-san, Sawada-san, 
and myself would always sit on the beautiful seashore and the younger 
ones listened to the physics discussion exchanged between Kikkawa-san 
and Sugawara-san. The one month stay in Honolulu was one of my great and 
most stimulating experiences.

At Kikkawa-san's 60th birthday Conference, everybody was talking about 
"p-branes and duality transformations".  I really thought we were 
timeslipping to 15$\sim$20 years ago.  At that time "dual transformation, 
membrane and n-dimensionally extended objects (now called p-branes)" 
were my favorite themes.~\cite{dual transform} ~\cite{membrane} If my 
paper on the membranes (which was the theory of n-branes) gave a little 
influence on the famous membrane paper by Kikkawa-san and 
Yamasaki-san~\cite{K-Y}, I would be very happy.
After moving to Ochanomizu University in 1987 from KEK, I have been 
working with my students mainly on the phenomenological problems of the 
non-Kikkawa type physics, including beyond the standard model effects in 
the $e^{+} e^{-}\rightarrow W^{+} W^{-}$ process, the effect of the top 
condensation in B-physics, the neutrino physics, the CP violating models 
and the baryogenesis of our universe.  Postdocs, Yasuhiko Katsuki, Kiusau 
Teshima, Hirofumi Yamada, Isamu Watanabe, Mohammad Ahmady and 
Noriyuki Oshimo did their own physics on the beyond the standard model, 
multiple production in perturbative QCD, non-perturbative QCD, linear 
collider physics and the two photon process, rare decays and the heavy 
quark symmetry in B-physics and CP violation in SUSY and SUSY breaking, 
respectively, with the help of the then students, Miho Marui, Kumiko 
Kimura, Atsuko Nitta, Azusa Yamaguchi, Fumiko Kanakubo, Tomomi Saito, 
Tomoko Uesugi, Tomoko Kadoyoshi, Minako Kitahara and Rika Endo.
I have, however,  sometimes come back to the Kikkawa-type physics on 
the string,  membrane and gravity theories with my postdocs and my 
students: For example,\\
(1) Orbifold models were firstly studied with Ikuo 
Senda.~\cite{orbifold}\\
(2) Using the light-cone gauge field theory of strings invented by Kaku 
and Kikkawa, we with Miho Marui and Ichiro Oda have derived the 
Altarelli-Parisi like evolution equation, since the decay function of 
strings works naturally in this light-cone frame as has happened similarly 
in QCD.~\cite{evolution equation} \\
(3) Knotting of the membrane was studied.~\cite{knot}\\
(4) With Ichiro Oda,  Akika Nakamichi and Fujie Nagamori, we studied 
four dimensional topological gravities, mainly on their 
quantization.~\cite{topological gravity} \\   
(5) Relating to this topological nature at high energies, estimation of the 
membrane scattering amplitudes is performed with Sachiko Kokubo, giving 
an indication of the structural phase transition among the intermediate 
shapes of the membranes, when the scattering angle is 
changed.~\cite{membrane scattering}\\
Other Kikkawa-type physics performed by our postdocs at Ochanomizu 
were;\\
(6) Kiyoshi Shiraishi studied some 5 years ago BPS soliton and 
Born-Infeld theories as well as the finite temperature field theories, \\
(7) The dilatonic gravity and black holes were investigated by Ichiro Oda 
and Shin'ichi Nojiri, and\\
(8) Hybrid model of continuous and discrete theories are examined by 
Toshiyuki Kuruma. 

Recently I am very much interested, as for the Kikkawa-type physics, 
in\\
(9) the generation of the Einstein gravity from the topological 
theory~\cite{string condensation},\\
(10) the swimming of microorganisms viewed from string and membrane 
theories,~\cite{swimming} and\\
(11) phase transition dynamics viewed from the field theoretical 
membrane theories.\\ 
The issue (9) is being investigated with Miyuki Katsuki, Hiroto Kubotani 
and Shin'ichi Nojiri, the issue (10) is with Masako Kawamura and Shin'ichi 
Nojiri and is helped by the Barcelona friends, Sergei Odintsov and Emil 
Elizalde, but the last issue (11) is still at the stage of promoting a vague 
idea.

In the next section I will mainly explain the issue (9), and will comment 
on my vague idea of the issue (11).
\section{Generation of the Einstein Gravity from the Topological 2-Form 
Gravity } 
The topological 2-form gravity is given by the following chiral action for 
the self-dual part:

\begin{equation}
  S = \int \frac{1}{2} \epsilon^{\mu\nu\lambda\rho}
\left( B^a_{\mu\nu}(x)R^a_{\lambda\rho}(x) + 
\underbrace{ \phi^{ab}(x)B^a_{\mu\nu}(x)B^b_{\lambda\rho}(x)}_
{constraint\: term} \right),\label{chiral action}
\end{equation}

where $B^a_{\mu\nu}(x)$ is the anti-symmetric tensor field or the 
Kalb-Ramond field and $R^a_{\lambda\rho}(x)$ is the $SU(2)$ field strengh 
for the $SU(2)$ spin connection ${\omega}_{\mu}^{a}$.  The constraint 
condition expressed by the Lagrange multiplier field ${\phi}^{ab}(x)$ can be 
solved naturally by introducing the vierbein and the t' Hooft symbol as
 $B_{\mu\nu}^{a} = \frac{1}{2} \eta_{BC}^{a} e_{\mu}^{B}e_{\nu}^{C}$.  Then 
we have the Einstein action.  In the process of solving the constraint the 
extra Kalb-Ramond symmetry  possessed by the topological "BF" theory is 
broken in an ad hoc way.   Instead we wish to start with the Kalb-Ramond 
invariant action and derive the constraint spontaneously.  The Kalb-Ramond 
symmetry, $B^a_{\mu\nu} \rightarrow 
B^a_{\mu\nu}+\nabla^{ab}_{\mu}\Lambda^b_{\nu} -\nabla^{ab}_{\nu}
\Lambda^b_{\mu}$, was originally the gauge symmetry of strings.  
Therefore, by introducing the string field, we write down the following 
Kalb-Ramond invariant action:

\begin{eqnarray}
  S&=& \int d^4 x\frac{1}{2} \epsilon^{\mu\nu\lambda\rho}
B^a_{\mu\nu}(x)R^a_{\lambda\rho}(x) \nonumber \\ 
& & + \sum_C \sum_{x_0 (\in C)} \sum_{x( \in C )}
 \epsilon^{\mu\nu\lambda\rho}
\left[ \left( \frac{\delta}{\delta C^{\mu\nu}(x)} + 
T^a B^a_{\mu\nu}[C; x, x_0]\right) \Psi[C; x_0] \right]
 ^{\dagger}\nonumber \\
& &~~~\times \left[  \left( \frac{\delta}{\delta C^{\lambda\rho}(x)} + 
T^a B^a_{\lambda\rho}[C; x, x_0]\right) \Psi[C; x_0]  \right] \nonumber \\
& &+ \sum_{C}\sum_{ x_0}  V [ \Psi[C; x_0] ^{\dagger} \Psi[C; 
x_0]] .
\label{ non-Abellian K-R action } 
\end{eqnarray}
In this expression we need to modify the Kalb-Ramond field
 $B^a_{\mu\nu}$ and its transformation to the non-local ones, relfecting 
the difficulty of their non-Abelian versions.  Now the condensation of the 
string fields 
\begin{equation}
 \frac{1}{2}\phi^{ab}[C; x_0] \equiv \langle\Psi[C; x_0] ^{\dagger} T^a 
T^b\Psi[C; x_0] \rangle,
\label{non-Abelian condensation}
\end{equation}
plays the role of the Lagrange multiplier.  If the condensation becomes 
large for the symmetric (isospin 2 ) part of $(a, b)$, then its coefficient 
gives the constraint,  leading to the Einstein gravity.  For the details refer 
to Ref.~\cite{string condensation}.
\section{Phase Transition Dynamics and Field Theory of  Membranes}
During the temporal development of the 1st order phase trandition, like 
the cooling down of the vapor (unbroken phase), liquid droplets of water
 (bubbles of the broken phase) are nucleated, they fuse with themselves, 
and finally the whole vessel (the whole space) is filled up with the water
 (broken phase).  It is really amazing to know that for such a difficult 
problem there exists a solvable theory called the Kolmogorov-Avrami 
theory~\cite{KA}, if the critical radius of the bubble is vanishing and the 
wall velocity is constant.  "Solvable" means that we can exactly know the 
probability of the arbitrarily chosen N spacetime points to belong to the 
broken or the unbroken phase.  This may suggest the existence of a solvable 
non-relativistic membranic (interfacial) field theory.  It is another 
Kikkawa-type physics to persue.



\begin{thebibliography}{99}
\bibitem{dual transform} A. Sugamoto, Phys. Rev. {\bf D19 } (1979) 1820 ;
 K. Seo, M. Okawa, and A. Sugamoto, Phys. Rev. {\bf D19 } (1979) 3744 ; K. 
Seo and M. Okawa, Phys. Rev. {\bf D21 } (1980) 1614 ; K. Seo and A. 
Sugamoto, Phys. Rev. {\bf D24 } (1981) 1630 .
\bibitem{membrane} A. Sugamoto, Nucl. Phys. {\bf B215 } (1983) 381.
\bibitem{K-Y} K. Kikkawa and M. Yamasaki, Prog.Theor. Phys. {\bf 76}
 (1986) 1379.
\bibitem{string condensation}M. Katsuki, H. Kubotani, A. Sugamoto and S. 
Nojiri, Mod. Phys. Lett. {\bf A29} (1995) 2143; M. Katsuki, S. Nojiri, and A. 
Sugamoto, OCHA-PP-61, NDA-PP-20 (1995), to be published in Int. J. Mod. 
Phys. {\bf A}.   
\bibitem{swimming} M. Kawamura, A. Sugamoto and S. Nojiri, Mod. Phys. 
Lett. {\bf A9} (1994) 1159 ; S. Nojiri, M. Kawamura and A. Sugamoto, Phys. 
Lett. {\bf B343} (1995) 181 ; S. Nojiri, M. Kawamura and A. Sugamoto, 
preprint, NDA-FP-21, OCHA-PP-65 (1995), to be published in Mod. Phys. 
Lett.{\bf A} ; E.Elizalde, S. D. Odintsov, S. Nojiri, M. Kawamura and A. 
Sugamoto, preprint, UB-ECM-PF 95/9-13, NDA-FP-22, OCHA-PP-66 (1995) 
hep-th/9511167; M. Kawamura, preprint, OCHA-PP-71, hep-th/9601156.
\bibitem{orbifold} I. Senda and A. Sugamoto, Nucl. Phys. {\bf B302}
 (1988) 291 ; Phys. Lett. {\bf 209B} (1988) 221 ; Phys. Lett. {\bf 211B}
 (1988) 308 . 
\bibitem{evolution equation} M. Marui, I. Oda and A. Sugamoto, Int. J. Mod. 
Phys. {\bf A5} (1990) 4257. 
\bibitem{topological gravity} I. Oda and A. Sugamoto, Phys. Lett. {\bf 
B266}, (1991) 280 ; A. Nakamichi, I. Oda and A. Sugamoto, Phys. Rev. {\bf 
D44} (1991) 3835 ; F. Nagamori, A. Sugamoto and I. Oda, Prog. Theor. Phys.
 {\bf 88} (1992) 797. 
\bibitem{membrane scattering} S. Kokubo and A. Sugamoto, Computer 
Phys. Comm. {\bf 75} (1993) 311 .
\bibitem{knot} A. Sugamoto, in the {\em "Proc. of the Trieste Conference 
on Supermembranes and Physics in 2+1 Dimensions}, M. J. Duff, C. N. Pope, 
and E. Sezgin (Eds.), World Scientific, pp16-28, (1990). 
\bibitem{KA}  A. N. Kolmogorov, Bull. Acd. Sci. U.S.S.R. , Phys. Ser. {\bf 3}
 (1937) 335 ; M. J. Avrami,   Chem. Phys. {\bf 7} (1939) 1103  ; S. Ohta, T. 
Ohta, and K. Kawasaki, Physica {\bf A140} (1987) 478 .
    
\end{thebibliography}
\end{document}